\documentclass[cits]{PoS}
\usepackage{graphicx}
\usepackage{amssymb}
\usepackage{fontenc}
\usepackage{times}
\usepackage{mathptmx}

\title{Distribution of High Mass X-ray Binaries in the Milky Way}

\ShortTitle{Distribution of High Mass X-ray Binaries in the Milky Way}

\author{\speaker{Alexis Coleiro}\thanks{We acknowledge A. Bodaghee, P.A. Charles, P.A. Curran, C. Knigge, F. Rahoui, M. Servillat and J.A. Zurita Heras for useful discussions. This work was supported by the Centre National d'Etudes Spatiales (CNES) based on observations obtained with MINE -- the Multi-wavelength INTEGRAL NEtwork--. This research has made use of the IGR Sources page maintained by J. Rodriguez \& A. Bodaghee (http://irfu.cea.fr/Sap/IGR-Sources/).}\\
        Laboratoire AIM (UMR-E 9005 CEA/DSM - CNRS - Universit\'e Paris Diderot),
Irfu / Service d'Astrophysique, CEA-Saclay, 91191 Gif-sur-Yvette Cedex, France.\\
        E-mail: \email{alexis.coleiro@cea.fr}}

\author{Sylvain Chaty\\
        Laboratoire AIM (UMR-E 9005 CEA/DSM - CNRS - Universit\'e Paris Diderot),
Irfu / Service d'Astrophysique, CEA-Saclay, 91191 Gif-sur-Yvette Cedex, France\\
Institut Universitaire de France, 103 boulevard Saint Michel, 75005 Paris, France\\
        E-mail: \email{chaty@cea.fr}}

\abstract{
Observations of the high energy sky, mainly with the \textit{INTEGRAL} satellite, have raised new questions about the formation and evolution of High Mass X-ray Binaries (HMXBs). The number of detected HMXBs of different types is now high enough to allow us to carry out a statistical analysis of their distribution in the Milky Way.
For the first time, we derive the distance and absorption of a sample of HMXBs using a Spectral Energy Distribution fitting procedure, and we examine the correlation with the distribution of Star Forming Complexes (SFCs) in the Galaxy. We show that HMXBs are clustered with SFCs with a typical cluster size of 0.3 $\pm$ 0.05 kpc and a characteristic distance between clusters of 1.7 $\pm$ 0.3 kpc. Furthermore, we present an investigation of the expected offset between the position of spiral arms and HMXBs, allowing us to constrain age and migration distance due to supernova kick for some sources. These new methods will allow us to assess the influence of the environment on these high energy objects with unprecedented reliability. }

\FullConference{"An INTEGRAL view of the high-energy sky (the first 10 years)"
9th INTEGRAL Workshop and celebration of the 10th anniversary of the launch,\\
		October 15-19, 2012\\
		Bibliotheque Nationale de France, Paris, France}

\begin{document}

\section{Introduction}

%High Mass X-ray Binaries (HMXBs) are binary systems composed of a compact object, a neutron star or a black hole candidate, accreting matter from a massive companion star: either a main sequence Be star or an evolved supergiant O or B star. Most of these sources are observed in the Galactic Plane \cite{Bird_2007} as it is expected for such young star systems which do not have time to move far from their birthplaces.\\
Thanks to the dedicated observations from \textit{RXTE} and \textit{INTEGRAL}, around 200 HMXBs are currently known in the Milky Way allowing us to focus on their distribution. Using RXTE data, \cite{Grimm_2002} highlighted clear signatures of the spiral structure in the spatial distribution of HMXBs. In the same way, \cite{Dean_2005}, \cite{Lutovinov_2005} and \cite{Bodaghee_2007} and \cite{Bodaghee_2012} showed that HMXBs observed with \textit{INTEGRAL} also seem to be associated with the spiral structure of the Galaxy. However, the HMXB positions, mostly derived from their X-ray luminosity, are not well constrained and highly uncertain in case of direct accretion as in HMXB. In order to overcome this caveat we present a novative approach allowing us to uniformly derive all HMXB positions. We study this distribution and the correlation with Star Forming Complexes (SFCs) observed in the Galaxy. Knowing the location of these sources, one can examine the composition of the environment at their birthplace. This study is necessary to better understand the formation and evolution of the whole population of HMXBs and primarily to state the role of the environment and binarity in the evolution of these high energy binary systems. For a complete overview of this study, the reader should consult \cite{Coleiro_2013}.
%In Section 2 we explain how we derive HMXB distances, then, in Section 3 we show that HMXBs are correlated with Star Forming Complexes. In Section 4, we derive the expected offset between HMXBs and Galactic spiral arms before discussing some implications on HMXB formation and evolution. We also derive the age and kick migration distance for 13 sources. Finally, we conclude in Section \ref{conclusion}.

%To compute HMXB distances, we gathered a sample of HMXBs for which at least 4 optical and/or near infrared (NIR) magnitudes were known. For each source we build its SED and fit it with a blackbody model. This enabled us to evaluate the distance of the source along with its associated uncertainty. 

\section{Deriving the HMXB location within our Galaxy}

\subsection{SED fitting procedure}\label{fit}

Using the Liu et al. catalogue (\cite{Liu_2006}), updated with literature, and the \textit{INTEGRAL} source catalogue of \cite{Bird_2010}, we retrieved the quiescent optical and NIR magnitudes (mostly from the 2MASS point source catalog), the spectral type and the luminosity class of each source from the literature. We then built a sample of 46 sources for which at least 4 optical/NIR magnitudes and the spectral type are known. For each source we build its optical/NIR SED and fit it with a blackbody model. Two parameters are fixed in the fitting procedure: the radius and the temperature of the companion star which dominates the optical and NIR flux. Two other parameters are left free: the extinction in V band $\rm{A}_{\rm{V}}$ and the ratio R/D, whereas the extinction $A_\lambda$ is derived at each wavelength from \cite{Cardelli_1989} assuming $R_{\rm{V}}=3.1$. Knowing the radius R of the companion star, we then calculate the distance D in kpc. Results on distance and extinction determination compared with previously published results are given in \cite{Coleiro_2013}. We point out that median discrepancy in distance is $\sim 17 \%$ and $\rm{A}_{\rm{V}}$ is often very similar (median discrepancy of less than 7\%).

\subsection{Uncertainties in the computed distance and extinction}\label{unc_section}

\begin{figure}[h!]
\begin{center}
\includegraphics[scale=0.3]{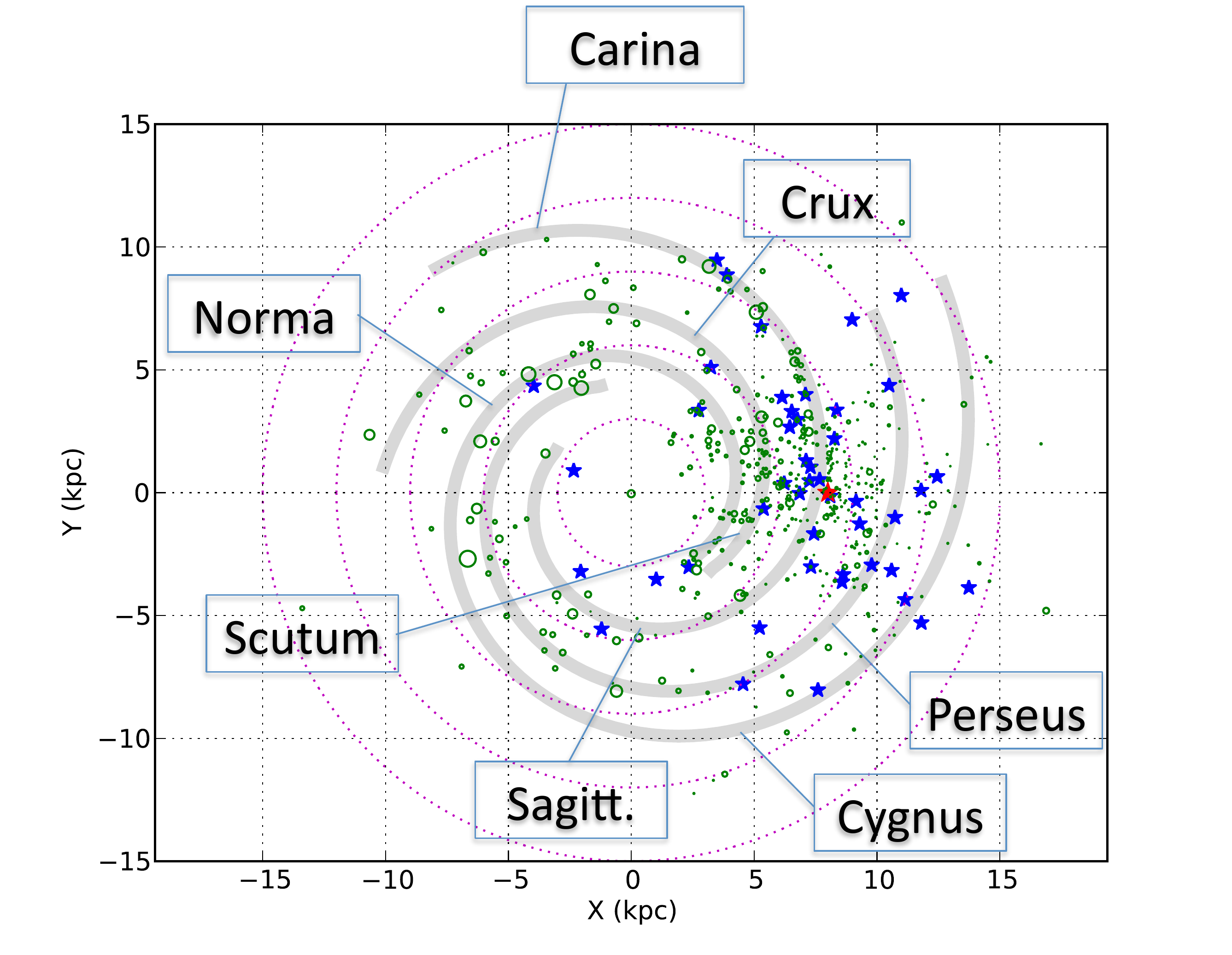}
\caption{Distribution of HMXBs (blues stars) and SFCs (green circles). The circle radius of SFCs represents the excitation parameter value. The spiral arm model from \cite{Russeil_2003} is also plotted and the red star at (8.5 ; 0) represents the Sun position.}
\label{disti}
\end{center}
\end{figure}

To compute accurate distance and extinction uncertainties, we estimated the magnitude uncertainties (retrieved from the literature), the uncertainties on the radius and temperature of the companion star, the degeneracy between several parameter values (based on the fitting procedure), the infrared excess of Be stars due to their circumstellar envelope and the error due to a different extinction law. For a complete analysis of these uncertainties see \cite{Coleiro_2013}. We present in Figure \ref{disti} the distribution of HMXBs in the Galaxy, obtained with our novative approach based on uniform distance determination. The spiral arms model given by \cite{Russeil_2003} is also presented. Figure \ref{errors} represents most of the studied HMXBs with the uncertainties on their location computed taking into account all the errors described above.
The question then arises: is there a correlation between this distribution of HMXBs and the distribution of Star Forming Complexes (SFCs) in the Milky Way (given by \cite{Russeil_2003}), as it is expected from the short HMXB lifetime ?  

\section{Results: HMXB distribution and correlation with Star Forming Complexes}\label{correlation_method}

%The first approach we adopt is to carry out a Kolmogorov-Smirnov test (KS-test) on each axis in order to quantify whether the two samples are drawn from the same probability distribution. We obtain a value of 0.15 for the X axis, a value of 0.25 for the Y axis and a value of 0.31 for the galactic longitude. These values are not negligible, suggesting that a correlation between the two samples does exist, though part of the information is lost because of the projection on the two axis. To overcome this caveat, we propose another method described hereafter.\\

To probe the correlation between HMXBs distribution and Star Forming Complexes, we use the method described in \cite{Coleiro_2013}. A correlation is detected with two characteristic scales: the typical cluster size of 0.3 kpc and the typical distance between clusters of 1.7 kpc with uncertainties of 0.05 kpc and 0.3 kpc respectively. If we take into account the uncertainties in HMXB positions and in SFC positions (given in \cite{Russeil_2003}, median error of 0.25 kpc), the correlation still exists with the same cluster size and the same distance between clusters. \cite{Bodaghee_2012} mention that HMXBs and OB complexes are clustered for a cluster size $r < 1$ kpc. This upper limit, obtained with a different method, is consistent with our results.

\section{Implication of the correlation on HMXB formation and evolution}

\begin{figure}[h!]
\begin{center}
\includegraphics[scale=0.3]{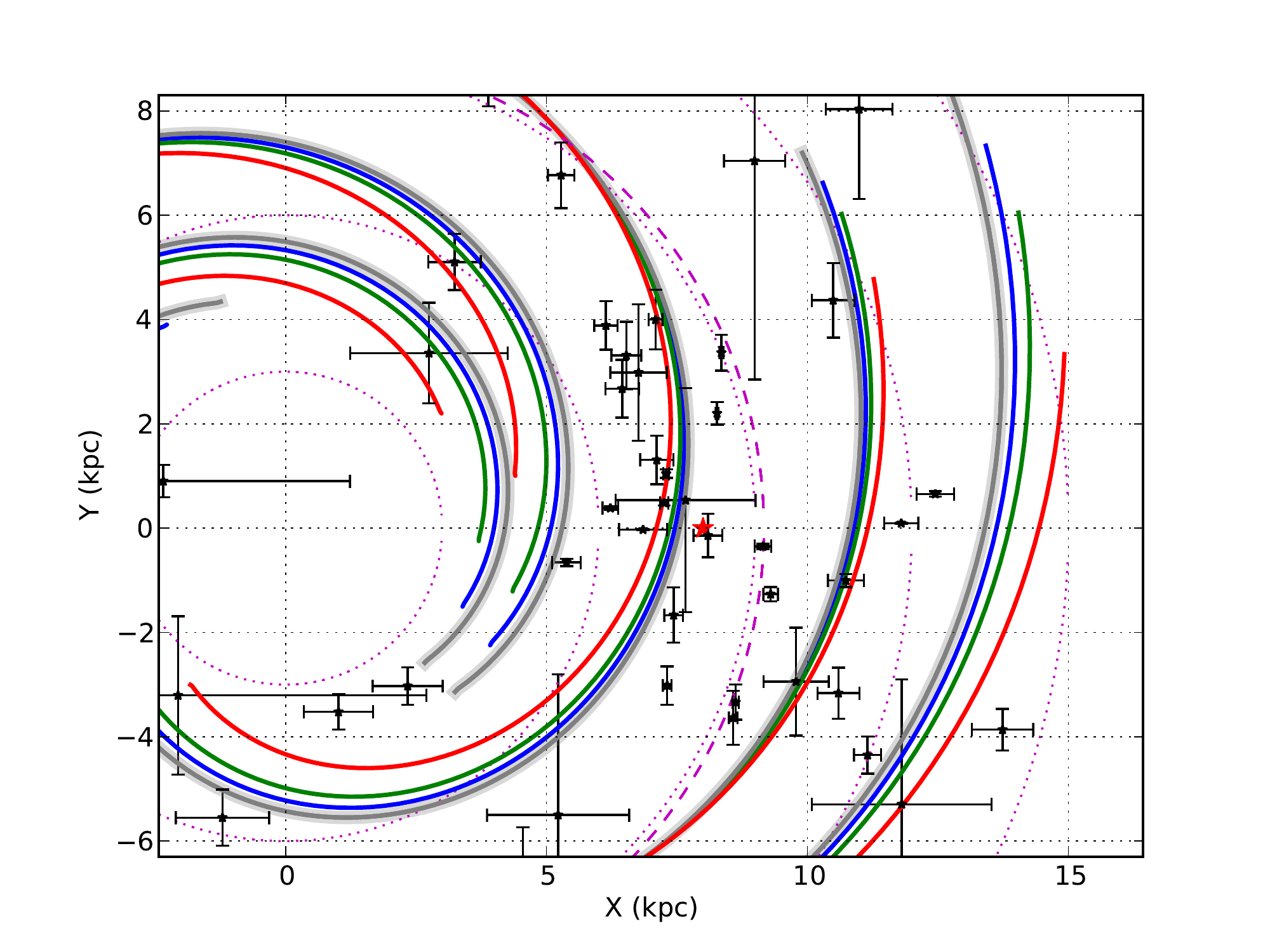}
\caption{Positions of HMXBs with error bars (zoom in the solar neighborhood). Color arms represent expected position of HMXBs with ages of 10 (blue), 20 (green) and 40 (red) Myr.}
\label{errors}
\end{center}
\end{figure}

The distribution of HMXBs reflects the stellar formation that took place some tens of Myr ago, since they are not an instantaneous star formation rate (SFR) indicator as explained in \cite{Shtykovskiy_2007}. Then, an offset between spiral arms (an indicator of the actual star formation) and HMXB distribution is expected. Indeed, since the spiral arm rotation velocity is different than the angular velocity of the stellar disk, we expect HMXB positions to be offset from the currently visible SFCs. This time lag was mentioned in \cite{Lutovinov_2005} and \cite{Dean_2005} but a deeper investigation of this issue was not possible due to the limited sample of HMXBs at that time. \cite{Shtykovskiy_2007} evaluated the offset for the galaxy M51. Here, we attempt to implement this formalism in the case of the Milky Way. One can assume the spiral arm density wave to rotate with the speed $\Omega_p$ = 24 km\,s$^{-1}$\,kpc$^{-1}$ (see e.g. \cite{Dias_2005}), in the same direction as the stellar disk, which velocity curve $\Omega(r)$ is assumed to be flat in the galactocentric distance range of interest, according to \cite{Brand_1993}. Then, to a first approximation, it is possible to derive the expected HMXB locations relative to the current position of the spiral arms in time $\tau$, i.e., the angular offset $\Delta\Theta(r)$ given by $\Delta\Theta(r) = (\Omega(r)-\Omega_p)\tau$. To estimate this displacement, we plot the expected positions of sources formed 10, 20, 40 Myr ago in the Galaxy map (Figure \ref{errors}). For better visibility, here, we do not plot the expected positions of sources formed 60, 80 and 100 Myr ago. Even if Figure \ref{errors} does not immediately suggest by eye any substantial offset between the current spiral arm position and the expected position of HMXBs (which depends on the age of the sources) we would now like to quantitatively assess the presence of an offset.

\subsection{Existence of an offset between HMXBs and Galactic spiral arms}\label{dist_calc}

To perform this study we must split the sample of HMXBs depending on the age of the sources. Two different samples are then created: one containing 4 supergiant stars and a second one containing 9 Be stars (the way these samples were created is explained in \cite{Coleiro_2013}). Then, we calculate the distance from each HMXB to the closest actual spiral arm (given in \cite{Russeil_2003}). We follow exactly the same procedure to calculate the distance from each source to the closest expected position of sources formed 20, 40, 60, 80 and 100 Myr ago and we determine the mean value of the offsets (taking into account all the sources of the two samples). The sources are expected to be closest to one of the expected positions computed above than to the current spiral arms observed by \cite{Russeil_2003}. We determine the 1-$\sigma$ error bars to the distance between each source $i$ and the closest point on the arm using the propagation of uncertainty formula. %The 1-$\sigma$ error associated to the distance between each source $i$ and the closest point on the arm is then given by the following equation:
Taking into account all the sources of the sample, we cannot observe any significant variation of the offset with time before 60 Myr. This result was expected because the minimum peak of distance between each HMXB and the closest expected position should be clearly different whether we consider Be or supergiant stars. Also, by taking all spectral types into account, we tend to loose a part of the information except the mean age upper limit of 60 Myr highlighted by the plot. For both supergiant and Be samples, we observe a more significant increase of the offset after 60 Myr. This increase could be a signature of the expected offset between the HMXB positions and the current spiral arms. However, we must be cautious about this result especially because of the small number of HMXBs in the two samples.

%Even if the locations of the sources are accurately determined, several reasons may affect the HMXB density and prevent the offset detection as underlined by \cite{Lutovinov_2005}: complex motion of density wave and stars from their birthdate to the X-ray phase, presence of previously undetected parts of the Galactic spiral arms, observational selection effect, etc. Finally, a larger sample of supergiant type HMXBs is needed to confirm the offset detection more confidently.\\

Moreover, the time interval during which HMXBs appear should translate the mass range of both stars of binary systems (see \cite{Dean_2005}). These results only enable us to state that this time interval is lower than 60 Myr on average for all stars.

\subsection{Deriving the age of HMXBs}\label{ageHMXBs}

Following the method described above, one can compute a distance between each source and the different theoretical arms that correspond to the current predicted position of a source sample born 20, 40, 60, 80 or 100 Myr ago. In the previous section we studied a sample of Be and supergiant HMXBs, and this could also be applied to each source separately, to constrain the age and the potential migration of the system due to a supernova kick.

%We choose not to take into account here the arm width that could be translated as an uncertainty on the position of the expected position of sources. Indeed, by taking this position dispersion into consideration, 1-$\sigma$ errors considerably increase and prevent any conclusion. Then, we assume all sources to be formed at the central position of the arm.\\ 

%We expect the distance from the source to the expected position to decrease until the expected position corresponds to the age of the source, and then to increase afterwards (see Figure \ref{distances_expl}).
It is possible to determine the rough age of these sources and a lower and upper limit of kick migration distances. If we consider only the sources studied separately here, we can derive a mean age of $\sim$ 45 Myr for sgHMXBs and 51 Myr for BeHMXBs. This result seems consistent with the distinct evolution timescales of these two kinds of systems but again, we must be cautious about this result because of the small sample of sources used in this calculation (4 supergiant and 9 Be systems).

\subsection{Constraints on migration from supernova kicks}

We focus on the sources selected before (4 sgHMXBs  and 9 BeHMXBs). The distance between the object and the closest expected position gives a kick value for each source and a mean value depending on the spectral type of the two samples. We derive a mean migration distance of 0.11 kpc for BeHMXBs and of 0.10 kpc for sgHMXBs. However, we should be careful about these mean values because of large error bars and small samples. It is also important to underline that these derived values only represent a lower limit to the kick migration distance since we cannot take into account the migration distance on galactic latitude given by the kick. Then, we only get the projected migration distance on the Galactic plane.

\section{Conclusion}\label{conclusion}
Examining the distribution of HMXBs is of major interest in order to study in depth the formation and evolution of these high energy sources. However, HMXB locations are usually poorly constrained and largely dependent on the determination method. Here, we determine the location of a sample of HMXBs using an uniform approach: SED fitting of their distance and absorption. This method enables us to reveal a consistent picture of the HMXB distribution, following the spiral arm structure of the Galaxy. This study shows that HMXBs are clustered with SFCs with a cluster size of 0.3 $\pm$ 0.05 kpc and a distance between clusters of 1.7 $\pm$ 0.3 kpc. We go further by quantitatively assessing the offset between current spiral density wave position and expected HMXB positions. Moreover, for 4 sgHMXBs and 9 BeHMXBs, we are able to derive an age and a migration distance, giving constraints on the supernova explosion kick.

\bibliographystyle{plain}
\bibliography{biblio_1}

\end{document}